\documentclass[12pt,preprint]{aastex}
\usepackage{natbib}

\begin{document}

\title{Nonthermal X-ray Emission in the N11 Superbubble in the 
Large Magellanic Cloud}

\author{L.A. Maddox, R.M. Williams\altaffilmark{1}, B.C. Dunne, and Y.-H. Chu}
\affil{Department of Astronomy, University of Illinois at Urbana-Champaign,
Urbana, IL  61801}
\altaffiltext{1}{Current Address: 
Columbus State University, Coca-Cola Space Science Center, Columbus, GA 31907-5645}

\begin{abstract}

We present the results of X-ray spectral analysis of the superbubble
around the OB association LH9
in the \ion{H}{2} complex N11 in the Large Magellanic Cloud. 
Using observations from {\em Suzaku}, we have modeled the X-ray emission
over the energy range 0.2-7.5 keV.  We contrained the thermal spectrum 
below 2 keV using a thermal plasma model found in a previous {\em XMM-Newton}
study of the N11 region.
We find that the hard X-ray emission ($>$ 2 keV) requires a nonthermal power-law
component.  The photon index of this component was found to be $\Gamma=1.72\pm 0.15$.
We performed an energy budget analysis for N11 using the known stellar
content of LH9.  We found that the observed thermal and kinetic energy in the 
superbubble is only half of the expected mechanical energy injected by stars.
\end{abstract}


\section{Introduction}

Superbubbles (SBs) are formed by the combined actions of stellar winds
from massive stars in OB associations and the eventual supernovae (SNe)
of those stars \citep{bruh80}. The physical structure of a SB is very 
similar to that of a bubble blown by the stellar wind of an isolated
massive star, as modeled by \citet{cast75} and \citet{weav77}, and
more recently by \citet{pitt01}.  The interior of a SB is shock-heated
by stellar winds and SN ejecta to $\sim10^6-10^8$ K and therefore
exhibits diffuse X-ray emission.  These regions of rarefied hot gas
reach diameters of $\sim100$ pc.  

The study of SBs in the Galaxy is impeded by the effects of obscuration
and confusion in the Galactic plane, making faint, extended objects such
as SBs difficult to observe, particularly at X-ray wavelengths.  In 
contrast, the high Galactic latitude and low internal column density of
the Large Magellanic Cloud (LMC) allows us a relatively un-obscured view
of its population of SBs, and as these SBs are all at roughly equal,
known distances \citep[50 kpc; ][]{feas99}, we can translate our 
observations into the physical properties of the SBs.

N11 is the second largest \ion{H}{2} region in the LMC after 30 Doradus, 
and  may constitute a more evolved version of this latter nebula 
\citep{walb92}. It harbors several associations of massive 
stars: LH9, LH10, LH13 and LH14 \citep{luck70}. Its structure 
is complex and reflects the interactions between the stars and their 
environment. 
The combined action of stellar winds and SN explosions from the central 
cluster LH9 has carved a hollow cavity in the surrounding ISM and
created a SB shell $\sim120$ pc in diameter.
The hot shocked winds/ejecta 
that fill this cavity emit X-rays  and provide 
the pressure to drive the SB shell expansion \citep{maclo98}. Only SNe
outside the SB can produce distinct SN remnants (SNRs) 
such as N11L at the western edge of the N11 complex \citep{will99}.

Previous studies of the N11 region with {\em ROSAT} and {\em XMM-Newton}
have presented the soft X-ray spectral properties of the SB
around LH9  \citep{maclo98,naze04}.
 Each study concluded that the soft
X-ray component was dominated by a thermal emission characterized by 
a temperature of $kT \sim 0.2$ keV. 
Investigations into the high-energy ($>2$ keV) properties of N11
were limited in both of these studies: {\em ROSAT} was not sensitive
to photon energies higher than $\sim 2$ keV; the {\em XMM-Newton} observations
suffered from
high particle background at these higher energies. 
The recently launched {\em Suzaku} has a moderate angular resolution, $\sim
2\arcmin$, and greater sensitivity to higher energy X-ray photons, providing
an excellent tool to study the high-energy diffuse emission in SBs.

In this paper we present the analysis of the X-ray spectrum of the 
SB around LH9 in N11 using new {\em Suzaku} observations.
In Section 2 we will provide a description of the observations and data processing procedures.
Section 3 will describe the spectral analysis of the main SB in N11 based on the {\em Suzaku} observations.
In Section 4 we will use the spectral results to perform
an energy budget analysis to compare the observed thermal 
and kinetic energies of the SB to the energy input based on the stellar population
of the OB association LH9. We will also discuss the possible mechanisms for the 
oberved non-thermal X-ray emission.

\section{Observations and Data Reduction}

The X-ray observatory {\em Suzaku} \citep[{\em ASTRO-E2};][]{mitsu07} 
includes
five X-ray telescopes (XRTs),
sensitive to soft X-rays up to $\sim10$ keV \citep{serl07}.
At the foci of four of the XRTs (XRT-I) are charge-coupled devices (CCD), known as
X-ray Imaging Spectrometers \citep[XIS;][]{koya07}.  Each CCD chip has an imaging area
of $1024\times1024$ pixels, with a corresponding pixels size of $24\mu\mathrm{m}\times
24\mu\mathrm{m}$. Three XIS CCDs (0,2 and 3) are front-illuminated (FI), and one (XIS1) is 
back-illuminated (BI).  The field-of-view for the XIS CCDs are  $18\arcmin\times18\arcmin$ with 
effective areas of 340 cm$^{2}$ (FI) and 390 cm$^{2}$ (BI) at 1.5 keV.
{\em Suzaku}
is located in a lower orbit than {\em XMM-Newton} and {\em Chandra}, providing a lower and more stable
particle background than other X-ray instruments \citep{mitsu07}.

N11 was observed with {\em Suzaku} on 2006 November 7-8 (ID 501091010; PI Williams) for a total
of 66.5 ks with a nominal pointing of R.A. 
$4^{\mathrm{h}}56^{\mathrm{m}}50\fs90$, 
Dec $-66\degr24\arcmin21\farcs6$ 
(J2000). Data were acquired on the three front-illuminated imaging
CCDs (XIS0, XIS2 and XIS3) and the one back-illuminated CCD (XIS1).
The data were processed with the HEASoft v6.6 software suite. We first applied
standard filters for bad pixels and event grades\footnote{As described in {\em Suzaku}
ABC Guide: http://heasarc.gsfc.nasa.gov/docs/suzaku/analysis/abc/}.
We removed time intervals when the source elevation was less than $10^{\circ}$
above the Earth's limb. Additionally, we removed times for when the satellite 
was above or within 436s of the South Atlantic Anomaly. 
Finally, we screened out events from the onboard $^{55}$Fe calibrator 
sources.
After all of the data screening, we 
obtained a final exposure time of 22.2 ks. 

For the XIS instrumental response, we  locally 
generated response matrices and ancillary response files using the tasks 
{\em xisrmfgen} and {\em xissimarfgen}, respectively.
The XIS background spectra were taken from a blank sky observation
toward the north ecliptic pole region, conducted for 95 ks on 2005
September 2--4.  Non-X-ray background spectra were taken from
night earth (NTE) observations compiled and available through the {\em Suzaku}
Science Team. The NTE spectra were weighted according to the 
magnetic cut-off rigidity of the source observation.
Source spectra were extracted using {\em XSELECT} from an 
elliptical region encircling the whole expanding SB around LH9.
The spectra were then binned with the task {\em grppha}
to include a minimum of 10 counts per bin over
the entire energy range.

An {\em XMM-Newton} observation of N11 was retrieved from the 
{\em XMM-Newton} data archive.  Analysis of this observation was
previously presented in \citet{naze04}.  Our data processing procedure
was identical to \citet{naze04}, and our spectral analysis results were consistent with
theirs.  We used the resulting thermal model to constrain the low energy spectrum
of the {\em Suzaku} observations. 

\begin{figure}
\begin{center}
\rotatebox{270}{\includegraphics[height=6.5in]{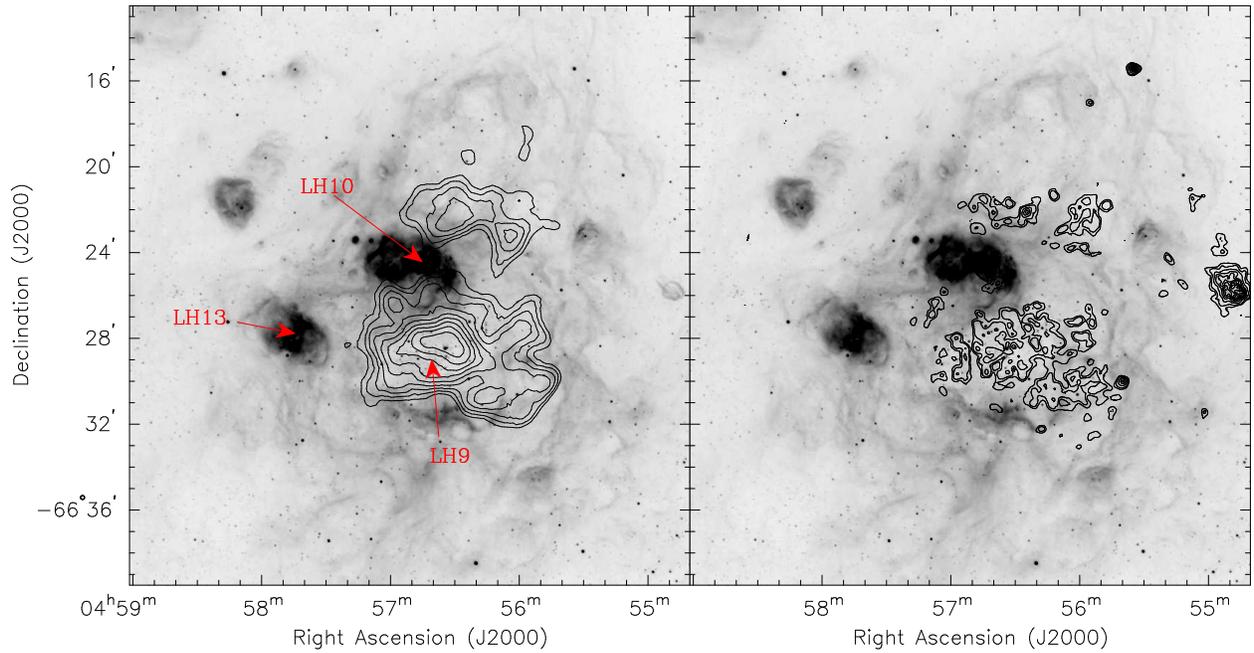}}
\end{center}
\caption{\label{cont} 
({\em left}) {\em Suzaku} soft (0.2--1.0 keV) X-ray contours overlaid on an 
H{$\alpha$} image of N11.  The
locations of the OB associations LH9, LH10 and LH13 are indicated by the arrows.  The
peak of the X-ray emission is coincident with the position of LH9.
({\em right}) {\em XMM-Newton} soft X-ray contours for the same region.
The SNR N11L to the right of the SB lies outside the {\em Suzaku} field-of-view.}
\end{figure}

Figure \ref{cont} shows the distribution of soft (0.2--1.0keV) X-ray emission
overlaid on an H$\alpha$ image of the N11 complex from the Magellanic Cloud Emission
Line Survey \citep[MCELS; ][]{smit99}.
The XIS contours were generated
using a combined image that was binned by a factor of 8 
 and smoothed with a 
Gaussian to the effective resolution of the {\em Suzaku} telescope 
$\sim2\arcmin$.  The peak soft X-ray emission in the {\em Suzaku} and 
{\em XMM-Newton} images are offset, but lie within the $\sim30\arcsec$ pointing
accuracy of {\em Suzaku}.  The emission peak lies within the evacuated
shell of LH9, where the stellar winds of the young stars have blown
away the natal gas that formed the cluster. The three strong point sources seen 
in hard X-rays (Figure 2) lie outside our spectral extraction region.

\section{Results}

Due to the difference in angular resolution between {\em Suzaku}
and {\em XMM-Newton}, 2$'$ and $\sim4\arcsec-6\arcsec$, respectively,
our extraction regions needed to be quite large.  We extracted
raw spectra from elliptical regions containing the entire SB around the OB association
LH9. The region measured $9\arcmin\times7\arcmin$
with its major axis aligned along the east-west direction.

The background subtraction in the {\em XMM-Newton} spectra yielded count
rates too low for meaningful fits above 2 keV, as was the case in \citet{naze04}.
For these spectra, we limited our fitting to the energy range 0.4--2 keV. 
The XIS spectra yielded a different challenge.  Above 8 keV, the subtraction
of the non-X-ray background was problematic, with incomplete subtraction
of Ni K$\alpha$ and K$\beta$ fluorescence lines.  This was not alleviated
with the subtraction of the exposure corrected blank-sky data.  For these
reasons, we limited our spectra fitting range to 0.2--7.5 keV.

We fit the {\em XMM-Newton} and {\em Suzaku} spectra separately using XSPEC v12.5.0
distributed with HEASoft 6.6.  The {\em XMM-Newton} data were used to contrain
the low energy ($E<2$ keV) thermal spectrum of N11.  We then used the resulting thermal
model as a fixed input to the {\em Suzaku} model fit. 
All fits include a fixed absorption column of 
$4.25\times10^{20}$ cm$^{-2}$ in the direction of the LMC \citep{dick90}.
An additional absorption component was used as a free parameter to account for the local LMC
column at the position of N11.
The first model fit attempted was a single {\em mekal} \citep{mewe85}
thermal plasma model, the model found best for the {\em XMM-Newton} spectra
in \citet{naze04}.  This single temperature model  was unable to fit 
the {\em Suzaku}
spectra above 2 keV.  We then added a second thermal component using various
models in order to fit the high-energy region of the spectra.  
 An additional thermal
component with a temperature of $kT = 0.8$ keV was able to improve the fit
at low energies ($E<2$ keV), but above 2 keV the models diverged
quickly from the observed high-energy spectrum.  
The only model 
that produced an improved fit, the non-equilibrium ionization model {\em nei},
yielded a gas temperature that was unphysical ($kT > 15$ keV) for a system of this age.  
We therefore ruled out a second thermal contribution to the spectra.

The best fit was achieved with a two-component model
consisting of a thermal plasma {\em (mekal)} and non-thermal power-law.
Initially, we allowed the metallicity to vary freely.  The resulting fits did
not deviate significantly from the average LMC abundances of 0.3 $Z_\odot$.
This is consistent with our understanding of this system as containing only
a small fraction of SN enriched material intermixed with the typical LMC ISM.
This also agrees with the results of self-enrichment models by \citet{nava08}, who
found that for OB associations with LMC-like initial conditions no measurable change
in metal abundances can be seen in 4-5 Myr, the age of LH9.
These current data are not sufficient to constrain abundance 
variations through SN enrichment over the broad spatial regions required by
the {\em Suzaku} spatial resolution.
We therefore fixed the abundance parameter to 0.3 $Z_\odot$ for the final model,
typical of the LMC ISM \citep{russ92}.
The total absorbed flux in the model is $(5.5\pm0.8)\times 10^{-13}$ erg s$^{-1}$ cm$^{-2}$.

The model indicates that the thermal component of the gas has a temperature
of $kT=0.18\pm 0.07$ keV, which is in good agreement with
the properties of other LMC superbubbles.  The error bars are 1-$\sigma$ uncertainty.
\citet{maclo98} obtained a temperature range of 0.1--0.2 keV to 99\%
confidence in the {\em ROSAT} observation.  The temperature
determination of \citet{naze04} was $kT=0.18^{+0.18}_{-0.17}$ keV. 

At higher energies ($>2$ keV), a non-thermal component is needed to fit
the X-ray spectrum. 
The photon-index ($E\propto\nu^{-\Gamma}$) of the best-fit model is 
$\Gamma = 1.72\pm0.15$, harder than would be expected for a synchrotron
mechanism.
The flux of the nonthermal emission is $(3.0\pm0.8)\times 10^{-13}$ erg s$^{-1}$ cm$^{-2}$, about 35\%
of the total flux. Figure \ref{spec} shows the final spectral model with the {\em Suzaku} and {\em
XMM-Newton} data. The best-fit parameters are shown in Table 1.

\section{Discussion}

We have used the observed stellar content of the OB association 
LH9 in the N11 SB to estimate the mechanical energy injected by 
stars via fast stellar winds and SN explosions.  We have also
used optical and radio observations to determine the kinetic
energy of the dense ionized and neutral gas in the SB, and X-ray
observations to assess the thermal energy of the hot interior 
gas. The comparison of these quantities were then used to determine
an energy budget for the SB. Finally we discuss the observed nonthermal
X-ray emission and possible physical mechanisms for this emission.

\subsection{Stellar Energy Input}

To estimate the stellar energy contribution for N11, we have 
considered the population of stars
in LH\,9 as given by \citet{1992AJ....103.1205P}.  In that 
study, bolometric magnitudes
and effective temperatures (as determined by their spectral 
types or $UBV$ colors) were used to determine stellar masses
and the present-day mass function (PDMF), then the PDMF is further
used to determine the initial mass function (IMF)
of the OB association.  \citet{1992AJ....103.1205P} noted
completeness problems for stars with $M \la 8\ M_\odot$.  
Ignoring the highest-mass bin due to potential evolutionary 
effects, \citeauthor{1992AJ....103.1205P} found 
the IMF to have a slope $\alpha= -$1.45$\pm$0.1 while 
\citet{1995ApJ...438..188M} found a IMF slope of 
$\alpha= -$1.38$\pm$0.16. We have adopted a middle value for 
the IMF of $\alpha= -$1.4.  

We have used the Starburst99 v5.1 package \citep{1999ApJS..123....3L} 
to model the stellar energy inputs of N11.
Starburst99 requires three basic inputs: an 
IMF slope, the total mass of stars formed
by the OB association in the 1--100 $M_\odot$ range (essentially 
a scaling factor for the IMF), 
and a timescale to integrate the
stellar energy inputs over.  
We use our adopted IMF slope of $\alpha=-1.4$, and the observed
main sequence star counts, to determine the total mass of
stars formed by the OB association LH9.
We note that the IMF will take the form 
$f(M) = KM^{\alpha -1} = KM^{-2.4}$, where $K$ is a constant that
can be determined from the star counts.  Using the star counts 
in the 12--15$M_\odot$, 15--20, 
20--25, and 25--40 $M_\odot$ bins from 
\citet{1992AJ....103.1205P}, we find $K=$3500$\pm$200.
The error bar reflects the uncertainty caused by integrating 
over different subsets of the total mass range.
To find the total mass of stars formed by the OB association, 
we integrated $f(M) dM$ over the 
1--100 $M_\odot$ mass range and find a total star mass of 
7400$\pm$400 $M_\odot$.  

To estimate a timescale, we
note that the earliest-type star observed in LH\,9 is an 
O6~V star, with a mass of 40-60~$M_\odot$.   
Such a star would have a main sequence lifetime of 
$\sim$4--5~Myr \citep{1993AAPS...98..523S}, 
and - assuming a single burst of star formation - gives us 
an estimate of the age of LH\,9. 
The Starburst99 model predicts a total wind input energy of 
8.6$\pm$0.5$\times$10$^{51}$~ergs at 5~Myr and
that 8$\pm$1 supernovae have already occurred, adding 
8$\pm$1$\times$10$^{51}$~ergs of
mechanical energy.  The total mechanical energy from LH\,9, 
as predicted from the star counts,
is therefore 17$\pm$2$\times$10$^{51}$~ergs. We assume that all of this
energy has gone into forming the SB.

\subsection{Observed Thermal and Kinetic Energy}

Using the best-fit model parameters, we are able to calculate 
some of the physical
conditions within the SB.  Since we used the {\em XMM-Newton} observations
to constrain the thermal spectral fits, we have included those data in our
calculations, leading to physical ranges. It should be noted that since our
extraction region is very large, the physical estimates will contain
hot gas that is not associated with LH9.  The figures given here should be
regarded as upper limits to the physical parameters within the main shell of
N11.

We can  estimate the electron density and mass of the thermal gas from the
normalization factor $A$ for the spectral fits:
\begin{eqnarray*}
A &\equiv& \frac{10^{-14}}{4\pi D^2} \int n_e n_{\mathrm{H}}\ dV,
\end{eqnarray*}
where $D$ is the distance to N11, $n_e$ is the electron density, $n_{\mathrm{H}}$ 
is the hydrogen density, and $V$ is the volume in cgs units. We assume that the 
hydrogen and helium is fully ionized, and that $n_{\mathrm{He}}/n_{\mathrm{H}}=0.1$; 
giving us a total particle density of $1.92n_e$. The electron density and hot 
gas mass are then calculated by the expressions
\begin{eqnarray*}
n_e &=& 3.89\times10^7 DA^{1/2}V^{-1/2}f^{-1/2}\ \ \ \mathrm{cm}^{-3},\\
M_{gas} &=& 1.17n_em_{\mathrm{H}}Vf\ \ \ \mathrm{g},
\end{eqnarray*}
 where $f$ is the
volume filling factor.
From these quantities, along with the plasma temperature, we can derive the 
thermal energy and pressure of the hot gas around LH\,9 by the following expressions:
\begin{eqnarray*}
E_{th} &=& 4.60\times10^{-9}n_e(kT)Vf\ \ \ \mathrm{erg},\\
P_{th} &=& 3.05\times10^{-9}n_e(kT)\ \ \ \mathrm{dyne\ cm^{-2}}.
\end{eqnarray*}

We calculate that the X-ray emitting gas has an electron density of
$n_e=0.098\pm0.003\, f^{-1/2}$
cm$^{-3}$, and that 
the total thermal energy for the gas is $2.3\pm 0.9\times 10^{51} f^{1/2}$ ergs.
The hot gas within the extraction region is calculated to have a total
mass of $2762\pm 85 f^{1/2}$ $M_\odot$.  Finally, the thermal pressure
of the emitting gas was calculated to be $5.38\pm 2.14\times10^{-11} f^{-1/2}$
dyne cm$^{-2}$.

It is most likely that the volume filling factor $f$ will range in value from 
$0.5-1.0$ due to the low density environment of the SB interior.
If we adopt a value of $f=0.75$ for the X-ray emitting gas, a medium value for 
the likely range, we calculate that the total observed 
thermal energy in the SB is $2.0\pm 0.8\times 10^{51}$ erg. 

\citet{rosa96} found that the N11 SB had an H$\alpha$ expansion velocity of 
45 km s$^{-1}$ in the shell. For a shell mass of $8.3\times10^4\ M_{\odot}$ 
\citep{dunn07}, we calculate that the \ion{H}{2} kinetic energy contribution
is $1.7\times10^{51}$ erg.
A survey of the \ion{H}{1} distribution of SBs in the LMC \citep{dunn07} found 
that the neutral gas bubble was expanding
at a rate of 25 km s$^{-1}$. With an \ion{H}{1} mass of $4.5\times10^5\ M_{\odot}$ 
we calculate a kinetic energy of $2.8\times10^{51}$ erg for the neutral shell.
So the total kinetic energy of the ionized and thermal gas is $4.5\times10^{51}$ erg, and the 
total observed energy (thermal plus kinetic) of the SB is $6.5\pm2.5\times10^{51}$ erg.

\subsection{Energy Budget}

The calculated thermal and kinetic energy ($6.5\pm 2.5\times 10^{51}$ erg) of the N11 SB 
represents an  energy deficit of about $62\pm13$\% when compared to the mechanical energy 
injected by the underlying OB association LH9 ($17\pm 2\times 10^{51}$ erg). 
This deficit is similar to  that found for N51D by \citet{coop04}. 
In that case, it was shown that the observed thermal and kinetic energy were 
about 3 times lower than the stellar energy input.  
Suggested energy loss mechanisms were a large scale blow out, where the hot gas breaks 
through the shell and escapes into the ambient ISM, and evaporation of dense cool gas 
into the hot interior of the SB.  Neither N51D nor N11 show evidence for blowouts, and
radiative losses due to the interface of cool and hot gas at the shell are not sufficient
to make up the energy deficit \citep{coop04}.

\subsection{Nonthermal X-ray Emission}

Our analysis of {\em Suzaku} observations of the SB N11 has demonstrated the
presence of a nonthermal high-energy component to the spectrum that was not
detected in previous {\em ROSAT} and {\em XMM-Newton} observations.  
This adds N11 to a small list of regions around OB associations that exhibit 
diffuse nonthermal X-ray emission: the SBs 30 Doradus C \citep{bamb04,smith04} 
and DEM L192 \citep[N51D,][]{coop04} in the LMC, and RCW 38 \citep{wolk06}
and Westerlund 1 \citep{muno06} in the Milky Way galaxy.  
Proposed explanations for the nonthermal emission in these objects are synchrotron 
radiation or inverse Compton scattering of starlight.  Each requires the acceleration
of electrons to relativistic energies.  Repeated SN shocks, turbulence in SNRs and colliding 
stellar winds, all expected in OB associations,  can accelerate the electrons to these energies
\citep{pari04}.

In the case of N11, the photon index ($\sim 1.72$) is harder than would be
expected for a aynchrtron origin.  It is more consistent with an inverse 
compton mechanism. 
The low angular resolution of the {\em Suzaku} observation makes it 
difficult to separate the individual contributions to the 
power-law emission, such as emission 
contributed by the stellar content of LH through the interaction
of the stellar winds with the surrounding \ion{H}{2} region, or SNR shock 
interactions with the SB shell. 
The raw {\em XMM-Newton} data indicate the presence of about six faint 
point sources within the spectral extraction region.
Four of these sources seem to have relatively hard X-ray colors, but
upon subtraction of background events, there are insufficient remaining photons
to do spectral analysis on the sources.  At least two of the {\em XMM-Newton} point sources 
contribute to the 1.0-2.0 keV emission along the NW rim of the SB.  In the
same region, one point source appears to contribute a few counts at energies $>$2 keV in the
{\em Suzaku} images.  The point sources contribute less than $\sim1\%$ of the total hard X-ray flux
that we observe in the SB X-ray emission, an insignificant fraction of the total.

The detection of nonthermal emission in N11, combined with previous 
SBs, demonstrates that faint nonthermal emission may be present in many SBs where
it has been undetected before due to limited high-energy sensitivity.
The discovery of a high-energy nonthermal component in N11 indicates that such 
components may not be uncommon in SBs; further observations with instruments 
such as {\em Suzaku} may be expected to increase the sample of such objects.  The 
origins of such components are not yet fully understood - and, indeed, multiple 
mechanisms may be at work.   If in general they are shown to be due to the 
presence of recent internal SNR shocks, as is indicated for N11, this will 
have interesting implications for the study of superbubble evolution and their 
role as cosmic-ray sources.
Due to the complex nature of SB environments,
it would be expected that nonthermal emission has multiple origins.
With a larger sample we will be able to study more closely the question of
what mechanism are generating this emission.

\section{Summary}

We have obtained {\it Suzaku} observations of the H II complex N11 in the LMC.
Prominent diffuse X-ray emission is detected from the central SB.
We have analyzed the {\it Suzaku} XIS data in conjunction with the {\it 
XMM-Newton} EPIC data.  We find a main thermal emission component
with a temperature of $kT \sim$ 0.18 keV, consistent with previous
analyses of {\it ROSAT} and {\it XMM-Newton} observations alone.
Owing to {\it Suzaku}'s high sensitivity at photon energies $>$ 3 keV,
we are able to extend the spectral analysis to hard X-rays and
find a nonthermal X-ray component with a photon index of $\Gamma \sim$ 1.72.
This is consistent with an inverse compton mechanism, though we cannot rule
out other mechanisms at this time.
This nonthermal component contributes $\sim$ 35\% of the total X-ray 
emission from the analyzed region.

We have performed an energy budget analysis for N11 using the known stellar
content of LH9.  The total stellar wind energy and supernova explosion energy
injected by LH 9 into the SB in the past 5 Myr are 
8.6$\pm$0.5$\times$10$^{51}$~ergs and 8$\pm$1$\times$10$^{51}$~ergs,
respectively.
The total thermal and kinetic energies retained in the N11 SB 
are 2.3$\pm$ 0.9 $\times$10$^{51}$~erg and 4.5$\times$10$^{51}$~erg,
respectively. 
We find a deficit of $\sim62\%$ in the thermal and kinetic energy stored in 
the superbubble from the expected mechanical energy injected by the stars.
This is similar to that observed in N51D. This could mean that these energy deficits
may be common in SBs. More sensitive observations of SBs will help answer this 
question.

\acknowledgements
The authors would like to thank the anonymous referee for helping to
improve the quality of this paper.
This work was supported in part by the following grants:
{\em Suzaku} grant NNX07AF61G, LTSA grant NNG05GC97G (RMW).

\clearpage


\begin{deluxetable}{lc}
\tablewidth{0pt}
\tablehead{\colhead{Model Parameter} &\colhead{LH9}  }
\startdata
Column Density $N_H$  ($10^{22}$ cm$^{-2}$) & $0.52\pm0.04$     \\
Plasma Temperature $kT$ (keV) & $0.18\pm0.07$    \\
Thermal Flux (absorbed) ($10^{-14}$ erg s$^{-1}$ cm$^{-2}$) & 25.4    \\
Thermal Flux (unabsorbed) ($10^{-14}$ erg s$^{-1}$ cm$^{-2}$)  & 954   \\
Photon Index ($\gamma$) & $1.72\pm0.15$   \\
Nonthermal Flux (absorbed) ($10^{-14}$ erg s$^{-1}$ cm$^{-2}$) & 29.6   \\
Nonthermal Flux (unabsorbed) ($10^{-14}$ erg s$^{-1}$ cm$^{-2}$) & 46.8   \\
$\chi^{2}$ (d.o.f)  & 544 (442)  \\
\enddata

\end{deluxetable}

\clearpage


\clearpage 

\begin{figure}
\begin{center}
\rotatebox{270}{
\includegraphics[height=6.5in]{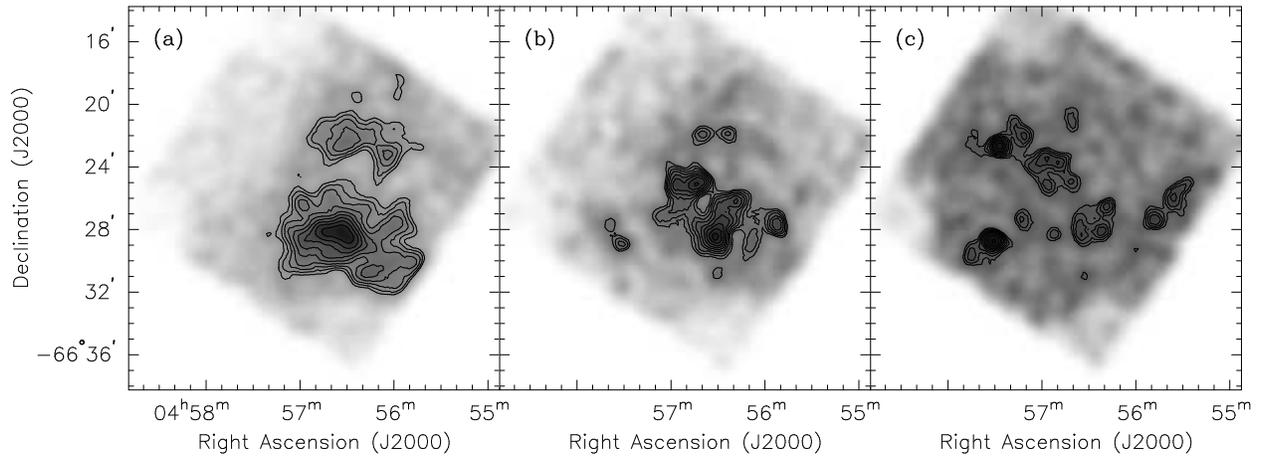}
}
\end{center}
\caption{\label{specext}
Combined XIS images of N11 with X-ray contours overlayed.  Energy ranges
for each image are (a) 0.2-1.0keV, (b) 1.0--2.0keV and (c) 2.0--10.0keV.}
%
%
\end{figure}

\clearpage

\begin{figure}
\begin{center}
\rotatebox{270}{\includegraphics*[height=5in]{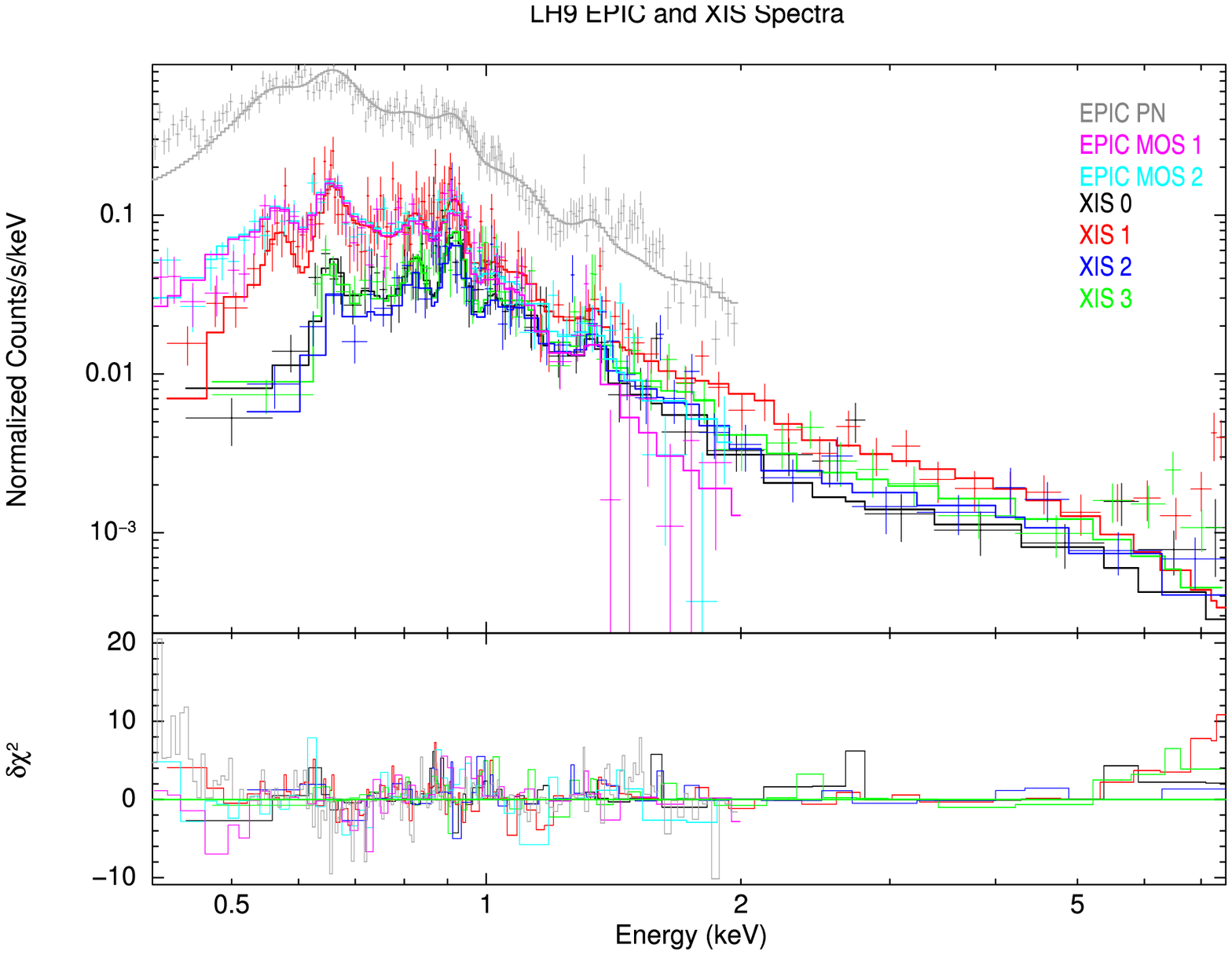},
\includegraphics[height=5in]{f3b.ps}}
\end{center}
\caption{\label{spec} ({\em top}) Combined spectra from all {\em XMM-Newton} and 
{\em Suzaku} detectors.  Colors denote the spectra from each instrument.
Solid lines show best-fit model. ({\em bottom}) 
Plot of the {\em Suzaku} spectra with the contributions from each model.
The spectral features below 2 keV 
represent emission line blends of highly ionized iron, neon and magnesium.}
\end{figure}

\end{document}